\documentclass{aa}  
\usepackage{natbib}
\bibpunct{(}{)}{;}{a}{}{,}

\usepackage{graphicx}
\usepackage{txfonts}
\usepackage{lipsum}
\usepackage{subcaption}         
\usepackage{lscape}         
\usepackage{placeins} 
\usepackage{booktabs}

\usepackage{hyperref}
\usepackage{amsmath}
\usepackage{makecell}
\usepackage{tabularx,xcolor}
\newcolumntype{Y}{>{\centering\arraybackslash}X}
\usepackage[para]{threeparttable}
\makeatletter
\def\TPT@dopara{\let\item\TPT@paraitem\par\noindent}
\makeatother

\makeatletter
\def\TPT@paraitem[#1]#2{\mbox{\textnormal{[\textit{#1}]}\,#2}\quad}
\makeatother

\begin{document}

\title{ALMA detection of continuum emission from GQ Lup B}

   \author{Pietro Pirola \inst{1}\corrauth{pietro.pirola@unimi.it}     
        \and Stefano Facchini \inst{1} \email{stefano.facchini@unimi.it}
        \and Gabriele Cugno \inst{2} \email{gabriele.cugno@uzh.ch}
        \and Myriam Benisty \inst{3} \email{benisty@mpia.de}
        \and Tomas Stolker \inst{4} \email{stolker@strw.leidenuniv.nl}
        \and Jaehan Bae \inst{5} \email{jbae@ufl.edu}
        \and Ruobing Dong \inst{6} \email{rbdong@gmail.com}
        \and Brendan P. Bowler \inst{7} \email{bpbowler@ucsb.edu}
        \and Ya-Lin Wu \inst{8} \email{yalinwu@ntnu.edu.tw}
        \and Richard Teague \inst{9} \email{rteague@mit.edu}
        \and Pietro Curone \inst{10} \email{pcurone@das.uchile.cl}
        \and Felipe Alarcon \inst{1} \email{felipe.alarcon@unimi.it}
        }
        
   \institute{Dipartimento di Fisica, Università degli Studi di Milano, Via Celoria 16, 20133 Milano, Italy 
   \and Department of Astrophysics, University of Zurich, Winterthurerstrasse 190, 8057 Zurich, Switzerland
   \and Max-Planck Institute for Astronomy (MPIA), Königstuhl 17, 69117 Heidelberg, Germany
   \and Leiden Observatory, Leiden University, Niels Bohrweg 2, 2333 CA Leiden, The Netherlands
   \and Department of Astronomy, University of Florida, Gainesville, FL 32611, USA
   \and Kavli Institute for Astronomy and Astrophysics, Peking University, Beijing 100871, People’s Republic of China
   \and Department of Physics, University of California, Santa Barbara, Santa Barbara, CA 93106, USA
   \and Department of Physics, National Taiwan Normal University, Taipei 116, Taiwan
   \and Department of Earth, Atmospheric, and Planetary Sciences, Massachusetts Institute of Technology, Cambridge, MA 02139, USA
   \and Departamento de Astronomía, Universidad de Chile, Camino El Observatorio 1515, Las Condes, Santiago, Chile
    }

   \date{Received 12-10-2026 / Accepted 05-11-2026 }

\abstract{GQ Lup B is a massive (10-40 M$_{\text{Jup}}$) companion orbiting the young T-Tauri star GQ Lup A at a projected distance of $\sim$110 au (0.7''). It is outside the circumprimary {disk in the submm continuum}, but its ${\sim}\,63^\circ$ inclined orbit intersects the primary's gas disk twice per period.
We report a 5.8$\sigma$ detection of continuum emission at the location of GQ Lup B, obtained with the Atacama Large Millimeter/submillimeter Array (ALMA) at 0.89 mm (band 7). We associate this emission with a dusty disk around the companion, already detected by VLT/NACO and JWST/MIRI in the mid-Infrared. To date, this is the third disk observed at mm wavelengths around {a companion spanning the brown dwarf and planetary mass regimes}, following PDS 70 c, a gas giant embedded in the parent circumstellar disk, and SR 12 c, a planetary-mass companion at 1200 au from the primary. 
We retrieve a point source with peak intensity $54.5\pm9.4\,\mu\rm Jy\, beam^{-1}$, co-spatial with GRAVITY astrometry. 
The emission is spatially unresolved, {placing} an upper limit on the source radius of 2.3 au {(from a synthesized beam of $\sim\,29\times25$ mas). For comparison, this is smaller than 1/3 of the Hill's radius (3-5 au for GQ Lup B, depending on the assumed mass), the theoretical upper limit for circum-planetary disks.} Assuming optically thin thermal emission from mm-size grains {in a disk-like structure}, we infer a mass of 0.0032 M$_\oplus$, while the optically thick limit yields an emission radius of 0.2 au. 
{Assuming a gas to dust ratio of 100, our results indicate a very low disk mass to accretion rate ratio, of the order of $\sim10^4$ years, analogously to the highest accretors in the stellar mass regime. In the case of GQ Lup B, this can be explained by its periodic crossings of the circumprimary disk, which could deliver a substantial amount of gas to the companion's disk over a long timescale.
}}


\maketitle
\nolinenumbers
\section{Introduction} 
{Typical tracers of an active planet formation process} are observed at optical and infrared wavelengths. UV excess, H$\alpha$ emission and near-IR hydrogen recombination lines trace ongoing accretion and shocks, \citep{zhou2014accretion,2021ApJ...917L..30A, demars2023emission} while continuum excess, silicate features in the mid-IR and dust-induced polarization are linked to $\mu$m-sized dust grains in the most irradiated surroundings of the protoplanet \citep{van2021survey,cugno2024mid, choksi2025spectral}. 

Observing these tracers requires high angular resolution and contrast, hence the most accessible targets are massive, bright and well-separated planetary mass companions (PMCs, \citealt{lachapelle2015characterization, bowler2016imaging}). 
These are gravitationally bound objects with masses broadly in the range 1–30 $M_\mathrm{Jup}$, at separations ${\gtrsim} 100$ au from the star and outside the primary disk.
{Despite forming via pathways more characteristic of
stars, their near-planetary masses suggest that they share properties with giant planets, including temperatures, surface gravities, and chemical compositions.}
{PMCs also represent an excellent laboratory to investigate the physics of disks around substellar mass objects. Disks around planetary-mass companions could provide a valuable and more accessible comparison to Circum-Planetary Disks (CPDs) \citep{canup2002formation,ward2010circumplanetary, keith2014accretion, szulagyi2016circumplanetary,maeda2024delivery} around planets embedded in the primary disks, helping us to understand their physical properties and evolution, and how they affect planetary formation mechanisms.}

For many PMCs rich observational datasets already exist, spanning direct imaging, accretion rate measurements and infrared spectral energy distributions (SEDs), {that could reveal the presence of a disk.} \citep{santamaria2018accretion,bowler2020strong,martinez2022mid, cugno2024mid,cugno2025_ctcha,2026A&A...710A.328D, 2026AJ....172..134F}. 
However, none of these observations are suitable for {investigating the bulk of a disk dust component}, as the emission is likely optically thick and only probes the hottest layers of the disk, those near the surface and close to the central object. 
The main component of the dust reservoir instead is likely located in the {colder region of the disk, in the midplane. It can be probed only by (sub)mm observations, which however may include contribution from free-free emission \citep{dominguez2025multi}.}
Detecting disks around PMOs and CPDs at millimeter wavelengths is extremely challenging and rare, as their typical dust mass seems to be lower than what can be probed by typical sensitivities of current surveys. 
Despite several ALMA campaigns targeting PMCs and embedded protoplanets
\citep{bowler2015alma, wu2017explanation,wu2020alma,jorquera2021search,facchini20262}, only two sources have been detected: PDS 70c \citep{isella2019detection,benisty2021circumplanetary}, inside the primary's disk, and SR 12 c \citep{wu2022alma}, one of the widest-orbit PMCs known to date. 
With this paper we present the detection of 0.89 mm continuum emission at the location of the {massive} companion GQ Lup B.

GQ Lup is located in the Lupus star forming region, at a distance of 154.1 pc  \citep{collaboration2022gaia}. GQ Lup A is a T-Tauri star with spectral type K7V, an age of 2-5 Myr and a mass of 1.03 M$_\odot$ \citep{donati2012magnetometry,macgregor2017alma}. 
It hosts a circumstellar disk, whose dust component extends up to {$59\pm12$} au, with an inclination with respect to the line of sight of $60.5\pm 0.5 ^\circ$ \citep{macgregor2017alma}. Previous ALMA observations show signs of a dust gap at ${\sim}10$ au from the star \citep{long2020hints} and an asymmetric CO disk ${\sim}\,200$ au wide \citep{macgregor2017alma}. Scattered light images with VLT-SPHERE show a single spiral arm in the southern region of the disk \citep{van2021survey}, likely caused by the interaction with the massive companion GQ Lup B. 
{This 10-40 M$_{\rm Jup}$ object} identified in 2005 \citep{neuhaeuser2005evidence}, is located at a projected separation of $0.7''$ from the primary. Recent studies, based on almost twenty years of astrometry and precise radial velocity measurements, found that its orbit is inclined by 63$^{+6}_{-14}{^\circ}$ with respect to the circumstellar disk, with a semimajor axis of 97 au \citep{venkatesan2025constraints}. {This orbital configuration suggests that} the companion punctures through the gas component of the primary disk {twice per orbit}, 
perturbing its kinematics and originating asymmetries in the disk morphology \citep{stolker2021characterizing}.
{Other than with the circumprimary disk, the companion's orbit is misaligned also with the stellar spin axis} (relative inclination $\sim\,52^{+19}_{-24}{^\circ}$ \citealt{bowler2023rotation,venkatesan2025constraints} ) 
possibly implying that GQ Lup B formed outside of the primary disk, from the same parent cloud of the primary star.
A similar {$^{12}\rm CO/^{13}\rm CO$ ratio} in the atmosphere of GQ Lup A ($51^{+10}_{-8}$) and B ($53^{+7}_{-6}$) \citep{picos2025eso} indeed suggests they originated from a shared material reservoir.
To complete the picture, there is a third object gravitationally bound to the system \citep{alcala20202mass}, GQ Lup C. It is a low mass star {(0.15 M$_{\odot}$)} at a projected distance of 2400 au, which possibly participates in the intricate dynamical architecture of the entire system and may have played a role in determining the highly inclined orbit of GQ Lup B. \\

The presence of a {disk} around GQ Lup B has long been discussed. Shortly after its discovery in 2005, hydrogen emission lines showed that the companion was still accreting material \citep{marois2007gq}. 
The hypothesis was further strengthened by the detection of UV excess \citep{zhou2014accretion}, tracing ongoing accretion, followed by mid-IR continuum excess which was explicitly linked with a disk-like structure \citep{stolker2021characterizing}. 
Recent JWST/MIRI observations confirmed that GQ Lup B's IR SED cannot be explained without accounting for the hot dust in a {disk around the companion}, with simple semi-analytical modeling suggesting a compact disk and moderately grown dust grains \citep{cugno2024mid}.
Previous ALMA observations set a 3$\sigma$ upper limit of 150 $\mu$Jy for ALMA band 7 continuum \citep{wu2017alma, macgregor2017alma}, leaving important disk parameters such as dust mass and structure unconstrained. 

\begin{table*}[t]
\centering
\caption{Summary of the observations used in this work.}
\begin{tabular}{cccccccc}
\toprule
ID & Name & Date & Baselines (L5 -- L80) & Frequency &  Number of EBs & PWV& On source time\\
 &      &      &     (m)     &   (GHz)         &      &    (mm)        &   (h)           \\

\toprule
2023.1.01340.S & wIB & 2024/07/13 & 99 -- 1003  & 332.81 -- 348.75 & 2 & 0.25 & 1.546\\
2024.1.00780.S & SB & 2024/10/03 & 33 -- 222   & 330.56 -- 345.82 & 1 & 0.89 & 0.672\\
2024.1.00780.S & ACA & 2025/04/07-10 & 9 -- 32     & 330.56 -- 345.82 & 3 & 0.91 & 1.739\\
2024.1.00780.S & IB & 2025/04/28-29 & 96 -- 1056  & 330.56 -- 345.82 & 3 & 0.91 & 2.167 \\
2024.1.00780.S & LB & 2025/06/10-16 & 208 -- 1915 & 330.46 -- 346.28 & 7 & 0.81 & 4.872\\
\toprule

\end{tabular}
\label{tab:observations}
\end{table*}

\section{Observations and data reduction}
We present new ALMA band 7 observations, combining two programs carried out between July 2024 and June 2025, targeting the GQ Lup system with 5 different baseline setups (ID 2023.1.01340.S and 2024.1.00780.S, PIs Ya-Lin Wu and Gabriele Cugno, respectively). In total, the observations consist of 16 Execution Blocks distributed over 13 nights. A summary of the MOUS (Member Observation Unit Sets) used in this work is presented in Table \ref{tab:observations}, including the baseline ranges, precipitable water vapor (PWV) at the time of the observation, frequency support and total integration time. {Throughout this paper,} different names will be used to refer to different MOUS: ACA refers to the data from the Atacama Compact Array; SB, IB, and LB refer to the Short, Intermediate, and Long baseline setup in program 2024.1.00780.S; wIB refers to the only MOUS of program 2023.1.01340.S.

Program 2023.1.01340.S was set up to optimize continuum detection. Its MOUS (wIB) consists of two Execution Blocks (EB) with four spectral windows each. Those spectral windows are centered in 333.787, 335.737, 347.736 and 345.782 GHz, with bandwidths of 1875, 1875, 1875 and 937.5 MHz respectively; the maximum spectral resolution is 0.565 MHz {($\sim 500$ m/s)}.
Program 2024.1.00780.S was tuned to optimize continuum detection and also study gas kinematics, as part of a joint ALMA-JWST proposal. 
Because of this, the ACA, SB and IB MOUS have two narrow spectral windows at high spectral resolution (a detailed discussion on gas kinematics will be the subject of a following work).
Their EBs have four spectral windows, centered in 330.567, 332.981, 343.627 and 345.774 GHz with nominal spectral resolution of 0.031, 0.977, 0.977, 0.031 MHz {($\sim0.03,0.9,0.9,0.03$ km/s)} and bandwidth of 58.59, 1875, 1875 and 58.59 MHz. 
LB MOUS were instead set to maximize continuum sensitivity: each one has four spectral windows centered in 331.400, 333.300, 343.450 and 345.350 GHz, with 1875 MHz bandwidth and 0.977 MHz spectral resolution.

\subsection{Self calibration}
After calibrating the raw data with CASA 6.6.1.17 pipeline (version 2024.1.0.8, CASA \cite{casa2022casa}), we performed self-calibration to improve the Signal to Noise Ratio (S/N) and increase chances of detecting the companion, adapting the procedure used by the exoALMA collaboration \citep{teague2025exoalma,loomis2025exoalma} to our specific combination of datasets.

Pipeline-calibrated measurement sets were first quality checked in amplitude-vs-channel and amplitude-vs-baseline space. No major outliers were found at this stage and consequently no data were rejected. Subsequently, a pseudo-continuum measurement set was created for each execution block individually by flagging channels containing spectral line emission ($^{12}$CO, $^{13}$CO, CS, SO) within $\pm$15 km/s of the systemic velocity and averaging the remaining data into 250 MHz channels. 
We checked for phase decoherence in individual EBs by looking for sudden decreases in amplitude against time \citep{loomis2025exoalma}, which were observed in most of our datasets. 
Therefore we performed one round of phase-only self-calibration on each EB, combining scans and spectral windows over the full observation duration. This reduced non-thermal noise and improved peak S/N by up to 50\%, providing cleaner input for the subsequent alignment step.
We then proceeded to combine different execution blocks into a single measurement set. Initially we combined all of the data from program 2024.1.00780.S, which share a common pointing center and frequency support. We then added the two EBs from program 2023.1.01340.S to the combined measurement set.

More specifically, starting with the most recent program, 2024.1.00780.S, EBs were aligned in the uv-plane by minimizing the weighted difference between visibilities gridded onto a common grid, considering only cells with overlapping coverage. This was performed by leveraging the script by \cite{loomis2025exoalma}.
We chose the best long baselines EB as reference (the one carried out on 2025/06/14, because of its almost circular beam and high S/N). Alignment then proceeded hierarchically: LB first, then IB referenced to the aligned LB, then SB to the aligned IB, then ACA to the aligned SB.

After spatial alignment, the self-calibration procedure was carried out starting from the shortest baselines EBs. We started by combining ACA and SB and performed self-calibration on the joint measurement set. Multiple phase-only rounds at decreasing solution intervals (360,180,120,60,30,18 s) were performed. To create the clean models used to calculate the gain tables we set a 6$\sigma$ threshold, masking an ellipse centered in the star's position with semimajor axis 0.7 arcsec, minor axis 0.35 arcsec and PA 346$^\circ$. 
After a full cycle of phase self-calibration, before adding longer baselines, flux offsets between EBs (from absolute calibration errors or baseline-dependent decoherence, choosing once again the highest S/N EB as reference) were corrected by rescaling the observed amplitudes. 
The whole phase self-calibration step was then started over with the rescaled EBs, checking that the flux offsets had been indeed corrected. 
The same process was then applied iteratively by progressively adding longer-baseline data (ACA+SB $\rightarrow$ ACA+SB+IB $\rightarrow$ ACA+SB+IB+LB), assessing phase decoherence and amplitude offsets at every step.  
After completing these rounds of phase-only self calibration, we applied two amplitude and phase self calibration rounds on EB-long and scan-long intervals.\\

The EBs from program 2023.1.01340.S were then added to the final measurement set with the same logic: first they were aligned in the uv plane to the visibilities obtained in the previous step, then phase only self-calibration was performed before rescaling fluxes; the process was started over on the rescaled EB and finally amplitude self-calibration was applied to the final measurement set. 
A 30s binning was finally applied to produce the definitive measurement set used in all the imaging procedures described in the next paragraph.

\subsection{Imaging}
\begin{figure}[t]
    \centering
    \includegraphics[width=1\columnwidth]{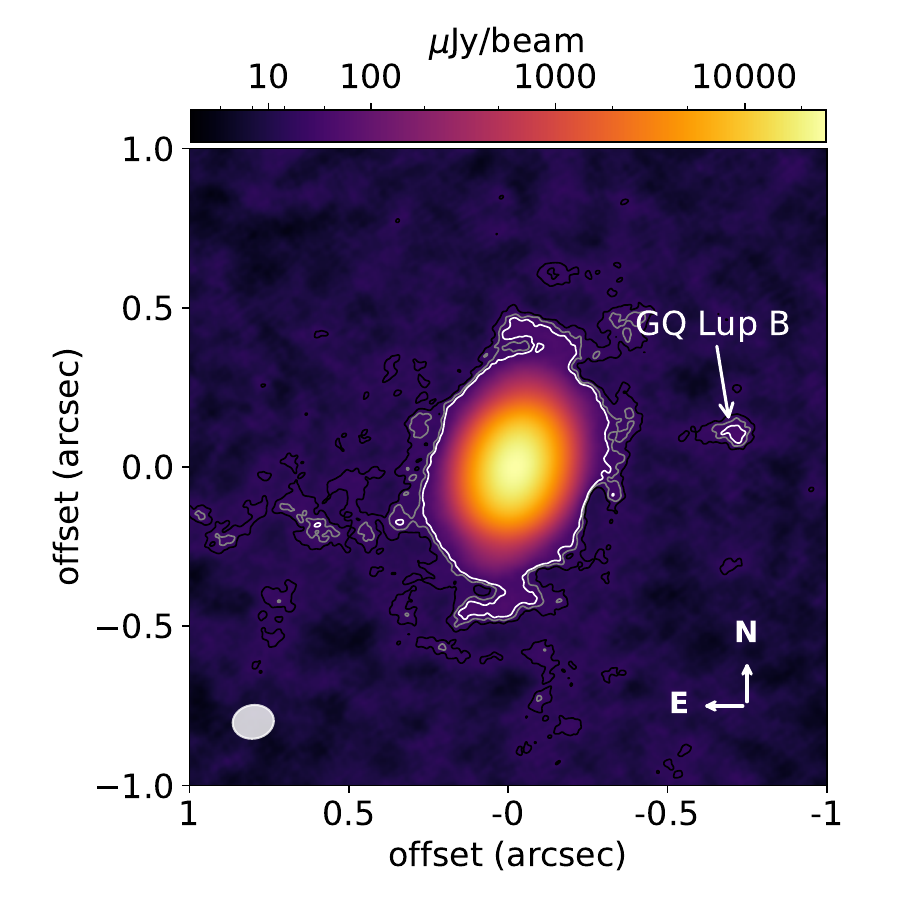}
    \caption{890$\,\mu$m continuum image of GQ Lup (robust=2, beam $\sim\,0.13 \times 0.11''$). The contours represent the 3,4, and 5$\sigma$ contours. Given the very high dynamic range of the image (${\sim}\,2900$), we used an Asinh normalization of the color scale.}
    \label{fig:continuum_fullscale}
\end{figure}

\begin{figure*}[t]
    \centering
    \includegraphics[width=1.\textwidth]{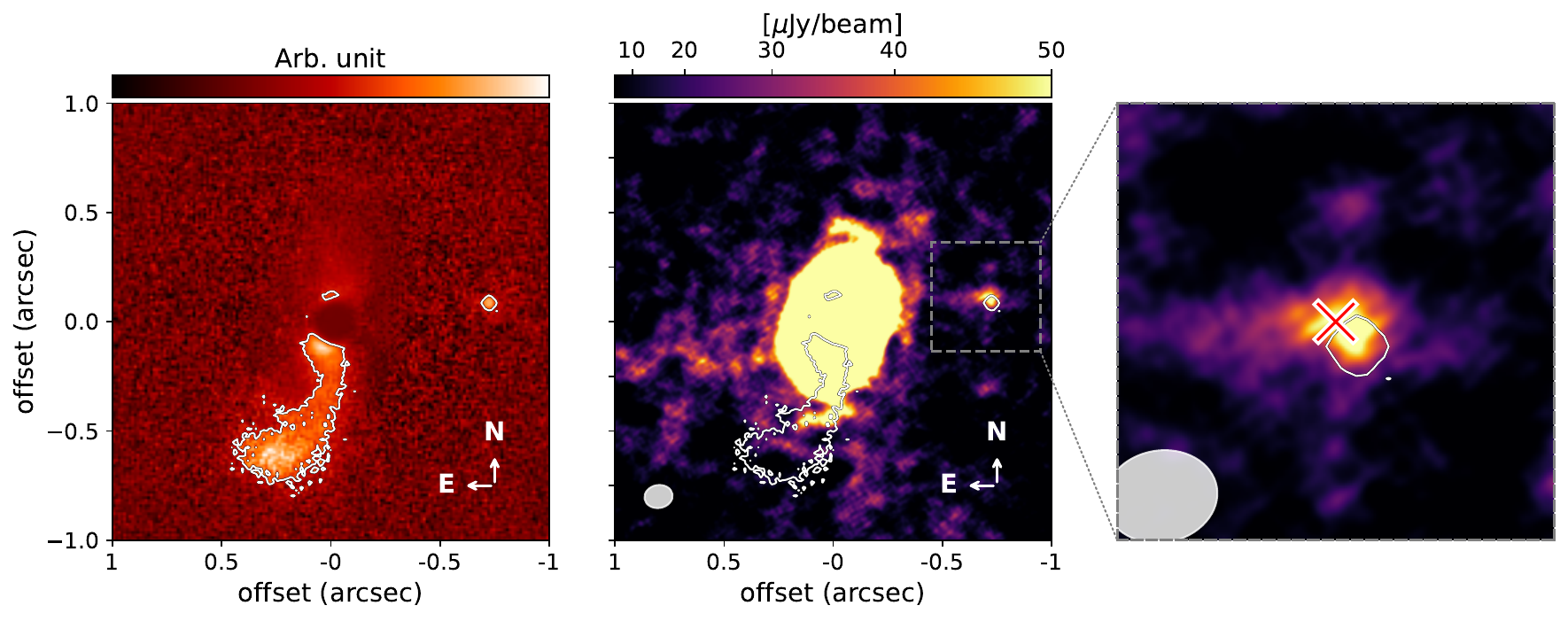}
    \caption{{\it Left:}  scattered light image of GQ Lup taken with VLT/SPHERE \citep{van2021survey}. A white contour (at 40\% of the peak of the image) serves as a reference for the other panels. The single spiral arm visible in the southern part of the image is attributed to the companion, which appears in the right side of the image. Although useful for spatial comparison, the apparent polarized flux of GQ Lup B is caused by the subtraction of the stellar component, as explained in the original paper \citep{van2021survey}.
    {\it Center:} saturated version of the continuum image from Fig. \ref{fig:continuum_fullscale}, where the color scale has been limited to a range of 0 to $50\,\mu\rm Jy \rm\, beam ^{-1}$. The dashed gray square, with sides of $0.5''$, is centered on GQ Lup B.
    {\it Right:} zoomed-in view of the same region as in the central panel: the red X marks the companion's position from GRAVITY {on 2023/3/8 \citep{venkatesan2025constraints}, while the white contour} represents the scattered light contour.}
    \label{fig:continuumvsscattered}
\end{figure*}

The final measurement set was imaged using the CASA v.6.5.4 tCLEAN algorithm. Given the very high dynamic range of the reconstructed images ($\gtrsim2900$), even minor bandwidth smearing can dominate the residuals away from the phase-center. We therefore used the Multi-Term Multi-Frequency Synthesis (mtmfs) deconvolver, with a first order polynomial (nterms = 2) and multi-frequency synthesis (mfs) spectral mode. Since we combined 7m with 12m MOUS, we used a mosaic gridder. 
To assign the clean components in the deconvolution algorithm, we masked the same ellipse used during self-calibration and a circle centered in the companion's position (radius $0.1''$, center RA offset $0.698''$, center DEC offset $0.114''$).
        
To assess the robustness of the structures we retrieve, we performed a grid of CLEANed images, using Briggs weighting with varying robust parameters. Depending on the robust parameter, our images have an rms ranging between 9.4 and 53.7 $\mu\rm Jy\,beam^{-1}$ with the corresponding beam size spanning from 0.13 to 0.02 arcsec. The rms was calculated on an emission-free region, inside an annulus extending between 2 and 6 arcsec from the pointing center.

\section{Results}

\subsection{Millimeter continuum emission from the GQ Lup system}
In Fig. \ref{fig:continuum_fullscale} we report the 890 $\mu$m continuum image we retrieve after combining all of our data.
{The primary disk is relatively bright, with a spatially integrated flux density of $70.11\pm 0.05$ mJy, calculated in the image plane with robust = 2 (over an elliptical area with PA = 347$^\circ$, semi-axes $0.4''$ and $0.5''$). This value is lower than the continuum flux reported in \citealt{macgregor2017alma} ($77.8\pm0.2\rm\, mJy$), obtained by fitting a two-dimensional gaussian to the continuum emission}.

Using the \texttt{galario} package, we fitted the visibilities modeling the disk with a central Gaussian with two external Gaussian rings, obtaining an inclination of $59.25 \pm 0.01 ^\circ$ and a position angle of $348.46 \pm0.02^\circ$, roughly consistent with the estimate obtained for the gas disk from the CO emission in \citealt{macgregor2017alma} ($i=60.5\pm0.5^\circ,\, \rm PA = 346\pm1^\circ$).
The disk appears fairly compact, extending up to $\sim 50$ au from the star. In Fig. \ref{fig:continuum_fullscale} we report 3,4, and 5$\sigma$ contours, which highlight the presence of extended, non axisymmetric emission in the northern and southern sides of the disk {at the 5$\sigma$ level}. This emission, {which remains visible even when varying the imaging parameters (see Fig. \ref{fig:varying_robust_combined}),} is not reported in previous ALMA observations{, which however had lower S/N and larger beams. The possible physical origin of these structures will be discussed in a forthcoming work, which will include different fitting procedures in the visibility plane.} 

Most notably, a point source is visible at the companion's location. The most significant detection ($5.8 \sigma$) is obtained with robust = 2, with a peak intensity of $54.5\,\mu\rm Jy\, beam ^{-1}$ and a 9.4 $\mu\rm Jy\, beam ^{-1}$ rms calculated in the same emission-free annulus described in the previous section. {We also calculate the rms in an annulus between 0.5'' and 0.8'' from the star, excluding a circle with 0.1'' radius centered in the companion's position. Over this area, the rms is 14.3 $\mu\rm Jy$, lowering the significance of the detection to 3.8 $\sigma$. We note, however, that this last annulus does not appear emission-free, as there is a faint filament visible at a $\sim3\sigma$ level in the east side of the disk (Fig. \ref{fig:continuumvsscattered}). This structure does not seem to be produced by imaging artifacts, as is apparent from the residuals images in Fig. \ref{fig:varying_thresh_combined}. We also note some structure on the east side of the companion's position, also at a $\sim3\sigma$ level.}

In Fig. \ref{fig:continuumvsscattered} we show a comparison with the position retrieved by GRAVITY astrometry \citep{venkatesan2025constraints} and with a scattered light image obtained with SPHERE \citep{van2021survey,stolker2021characterizing}. { The peak of the intensity in our ALMA image is $\sim\,0.02''$ from the position from GRAVITY reported in Fig. \ref{fig:continuumvsscattered} (2023/3/8) and $\sim 0.01''$ from the peak of the image from SPHERE. The latter is useful for spatial comparison, but as noted in the original article \citep{van2021survey}, GQ Lup B's emission is not polarized up to $\sim\,0.2\%$. The apparent signal is instead spuriously created by subtracting the stellar polarization. }
Since all observations were taken within a year from July 2024 to June 2025, and given the companion's orbital period of $\sim$ 1000 years \citep{stolker2021characterizing,venkatesan2025constraints}, we can safely assume that its position did not change significantly between different measurement sessions. Specifically, GRAVITY data collected between August 2021 and March 2023 \citep{venkatesan2025constraints} reveal a proper motion of GQ Lup B relative to its host star of ${\sim}\,  4$ mas in the same time span, five times smaller than the smallest synthesized beam we can obtain. We therefore ignore any proper motion in our analysis.
The coincidence of the point source seen with ALMA to both the IR signals suggests that their origin is co-spatial. We interpret this emission as thermal emission from millimeter-sized grains in an unresolved disk surrounding GQ Lup B. 

\begin{figure}[h]
    \centering
    \includegraphics[width=1\columnwidth]{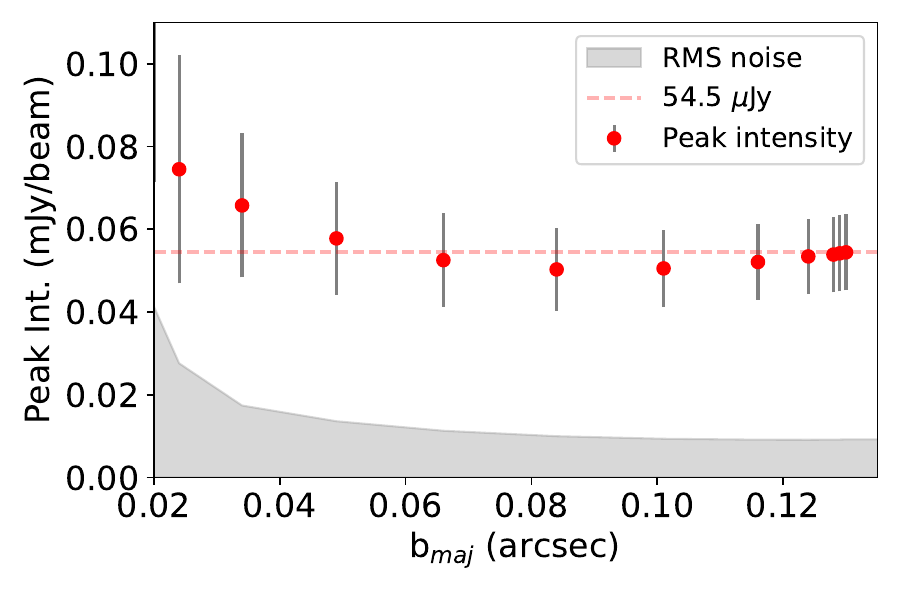}
    \caption{Peak intensity at the companion's location obtained with different Briggs robust parameters, which translate into different beam sizes.}
    \label{fig:test1}
\end{figure}

To verify that the emission is not due to imaging artifacts we carried out two sets of tests. 
Following \cite{benisty2021circumplanetary}, we produced multiple images varying the robust parameters (and consequently the beam sizes, {see Fig. \ref{fig:varying_robust_combined} in the appendix)}. 
If the emission was due to the contamination from GQ Lup A's disk, the peak intensity would increase with more extended beams, since the resolution element would include a higher contribution from the primary. This is not the case for GQ Lup B: regardless of the beam size the observed peak intensity remains stable within a 1$\sigma$ interval from the reported value of 54.5 $\mu$Jy beam$^{-1}$ (see Fig. \ref{fig:test1}). 
The constant peak intensity also indicates that the emission is unresolved independently of the imaging parameters. {Furthermore, Fig. \ref{fig:beam_radial_profiles} highlights how the radial profile of the emission behaves as a point source with peak intensity just above the cleaning threshold}.
The implications for the disk's size are discussed in the next section.  

The images we obtain have a very high dynamic range, with the peak intensity of the primary disk reaching up to 2900 times the rms noise level. When searching for faint structures, this must be treated with caution, as sidelobes from the bright primary could potentially be mistaken for real emission.
To address this, we produced an additional image set masking only the primary disk (see Figs. \ref{fig:test2} {and \ref{fig:varying_thresh_combined} in the appendix)}, preventing the cleaning algorithm from placing any CLEAN components at the planet's location. Since the algorithm iteratively subtracts model components convolved with the dirty beam, which is responsible for the sidelobes, any spurious emission arising from those sidelobes should be progressively reduced as the cleaning proceeds deeper. If the 5.8$\sigma$ point source was indeed an artifact, we would expect its flux density in the residual image to decrease with the clean stopping threshold. Instead, the flux density measured at the companion's location in the residual image remains constant, further indicating that the detected emission is not related to the primary's sidelobes.

\begin{figure}[h]
    \centering
    \includegraphics[width=1\columnwidth]{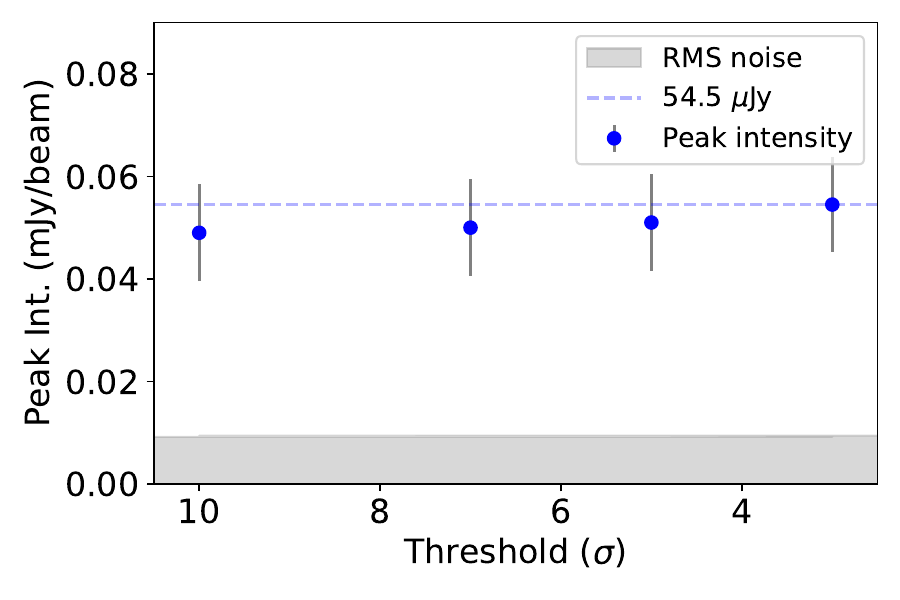}
    \caption{Peak intensity at the companion's location in the residual map, masking only the primary disk and stopping the imaging procedure at different thresholds.}
    \label{fig:test2}
\end{figure}

\section{Discussion}
Our detection represents the third observation of continuum emission from a bound planetary-mass object in the millimeter regime, the second around a wide-orbit PMC \citep{wu2022alma}. 
The mid-IR excess observed in the GQ Lup B's SED has previously been associated with a dusty disk \citep{cugno2024mid}. 
Therefore, it is intuitive to associate the emission we retrieve at 890$\,\mu$m with thermal emission from dust grains in a disk-like structure. {However, recent studies on PDS 70 c \citep{dominguez2025multi,casassus2026hourly}  highlighted that other mechanisms (e.g. free-free emission from accretion shocks) could also play a role at submm wavelengths.
Only a multi-band characterization and thorough radiative transfer modeling can distinguish between different scenarios. In the next sections, we discuss the implications of our single measurement within the assumption that the point source we detect is caused only by thermal emission from a dusty disk.}

We note once again that GQ Lup B's IR emission observed with SPHERE is not polarized. If we consider only emission from a disk, this could imply a low inclination \citep{van2021survey} or alternatively relatively grown dust grains ($a\gtrsim5\mu\rm m$, \citealt{cugno2024mid}). 

\subsection{Flux comparison with stars and PMOs}
\begin{figure*}[t]
    \centering
    \includegraphics[width=1\textwidth]{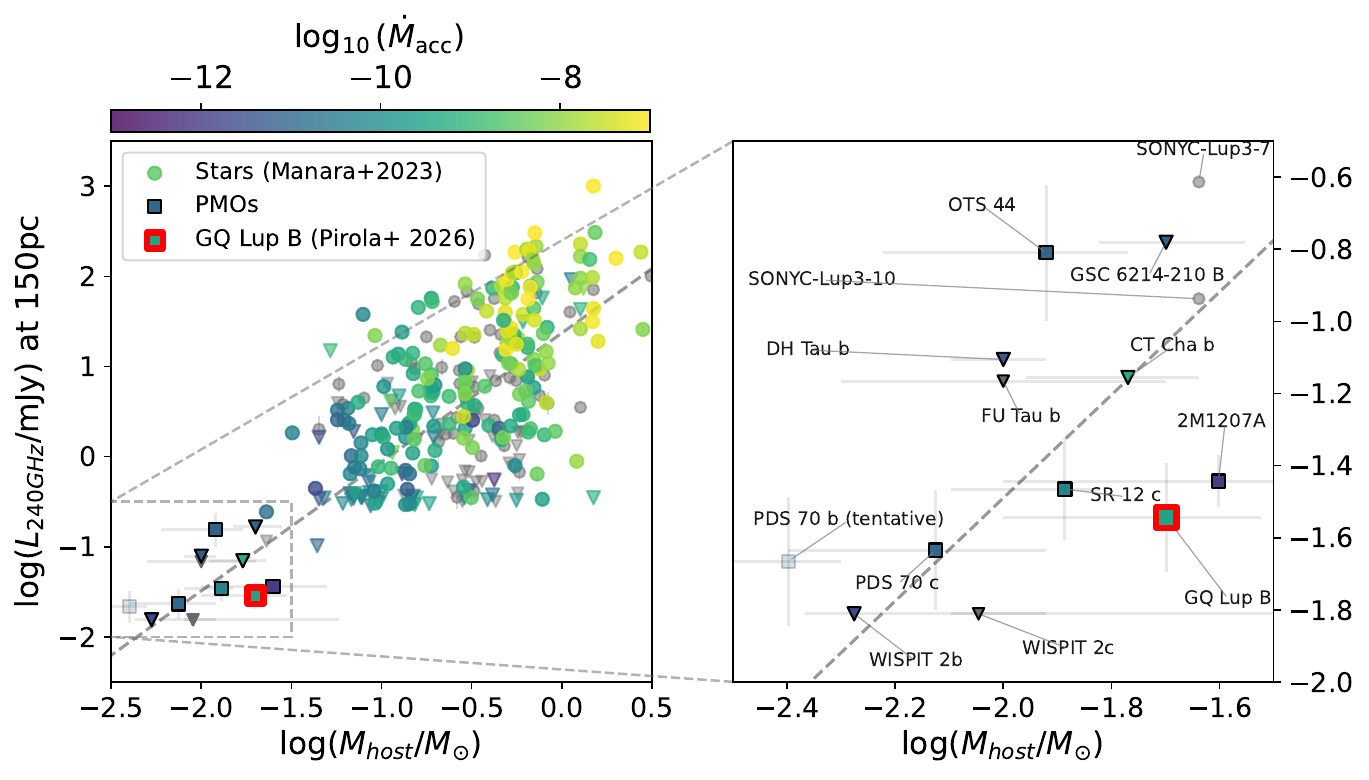}
    \caption{
    Measured fluxes and upper limits for stars and planetary-mass objects, rescaled to a common frequency and distance, as a function of the host mass. 
    Stars \citep{manara2023demographics} are reported as circles, while squares refer to planetary-mass objects. Other than PDS 70c and SR 12c, we report the free floating OTS 44 \citep{bayo2019modeling} and the low-mass brown dwarf 2MASS J12073346-3932539 A (2M1207A) \citep{ricci2017alma}. The faded dot represents the tentative detection of PDS 70 b \citep{benisty2021circumplanetary,fasano2025}. Triangles represent 5$\sigma$ upper limits for GSC 6214-210 b \citep{bowler2015alma}, DH Tau b, CT Cha b and FU Tau b \citep{wu2020alma}, and for the embedded planets WISPIT 2c and WISPIT 2b \citep{2026ApJ..1009L..32B,facchini20262}. 
    Red represents our measurement of GQ Lup B. {The color scale accounts for the host's accretion rate (see discussion in Sect. 4.3). Stars without an accretion rate measurement are colored in gray.} Disks around PMOs seem to broadly follow the relation found for stars, represented by the dashed gray line. For GQ Lup B we assume a mass of $20\,M_{\rm Jup}$ as a mean value between different estimates, which vary between 10 and 30 $M_{\rm Jup}$ \citep{marois2007gq,seifahrt2007near,stolker2021characterizing,cugno2024mid}}
    \label{fig:fvsmass}
\end{figure*}

In Fig. \ref{fig:fvsmass} we compare the submm flux density of GQ Lup B to other disks around stars and planetary mass objects, rescaled to a common frequency (240 GHz) and distance (150 pc). 
We adopt a spectral index of 2.2 for protoplanetary disks \citep{2020ARA&A..58..483A} and 2 for CPDs and disks around PMOs, as measured for PDS 70 c \citep{fasano2025, dominguez2025multi}.

In general, disks around stars seem to follow a common trend, with slope $\beta = 1.43 \pm0.10$ and intercept $\alpha = 1.37 \pm 0.06$ \citep{Andrews2021,manara2023demographics, facchini20262}, with a considerable scatter around the best fit relation. 
This large scatter, together with the scarce number of detections, makes it premature to assess whether also planetary-mass objects follow the same relation.
GQ Lup B exhibits lower peak intensity in ALMA band 7 than PDS 70c and SR 12c ($54.5\pm9.4$, $86\pm16$ and $127\pm14\,\mu\rm Jy\, beam^{-1}$, respectively), {but when rescaled to a common distance the measurements seem to align ($28.6\pm4.9$, $23.2\pm4.4$ and $34.3\pm3.8\,\mu\rm Jy\, beam^{-1}$, at 240 GHz if the sources were at 150 pc)}. These already available detections indicate the sensitivity threshold necessary to detect {disks around substellar objects at mm wavelengths}, increasing the sample needed to understand similarities and differences between circumstellar and circumplanetary physics.

\subsection{Mass and radius}

\paragraph{Constraints on the radial extent}
Since the emission around GQ Lup B is unresolved, only limited constraints on its size and mass can be derived. 
An agnostic estimate on the disk size can be given directly by {inspecting} the submm continuum images.
As shown in Fig. \ref{fig:test1}, the disk remains unresolved even at the highest angular resolution achievable with our dataset ($\sim0.03''$): if it was resolved, we would expect the peak intensity to drop as the angular resolution increases. Instead, the peak intensity remains constant across all the images independently of the synthesized beam size.
Varying the robust parameter, we can improve the angular resolution at the cost of increasing the noise. The smallest synthesized beam for which the signal-to-noise ratio exceeds 3 is $\sim29\times25$ mas, obtained with robust -0.5. At the system's distance of 154.1 pc, this corresponds to approximately 4.6 au. Therefore we adopt $\sim2.3$ au as a conservative upper limit on the outer radius {of the dust component of the disk.} 
This is slightly smaller than one third of the companion's Hill radius (3-5 au, when adopting the most recent orbital parameters from \citealt{venkatesan2025constraints}), which is usually cited as the theoretical upper limit for embedded CPDs \citep{martin2011tidal,shabram2013evolution}.
The upper limit we find for GQ Lup B is also comparable with the upper limit of 5 au for SR 12 c and 1.2 au for PDS 70c, confirming that independently of the separation from the primary, these disks are likely compact in their (sub)mm continuum emission.

Simulations in \cite{rab2019observing} showed that PMC disks can be optically thick at millimeter wavelengths. Under this assumption, we can obtain an estimate for the disk outer radius, since in this regime the observed flux density scales with the emitting surface. 
Following \cite{isella2014searching,isella2019detection,wu2022alma} we start by assuming a temperature profile for the disk's {dust}: 
\begin{equation}
    T(r)^4 = T_{*,irr}^4 + T_{\rm irr,p}^4 + T_{acc}^4.
    \label{eq:temperature}
\end{equation} 
The first component is an approximation of the stellar contribution, which we compute as
\begin{equation}
    T_{\rm irr,*}^4(R) = \frac{\phi L_*}{8\pi R^2 \sigma_{\rm SB}}.
\end{equation}
This is the simplified expression for a passively heated, flared disk in radiative equilibrium \citep{chiang1997spectral,d1998accretion,dullemond2001passive}, where $L_*$ is the stellar bolometric luminosity and $\sigma_{\rm SB}$ is the Stefan-Boltzmann constant. In the original papers $\phi$ is the flaring angle of the disk surface: in this case it represents the fraction of stellar radiation that hits the disk surface due to its inclination, and we take it as the sin of the grazing angle between the irradiation and the disk plane; $d$ is the distance between the companion and the star. 
Also the companion contributes {to the total temperature profile }\citep{isella2019detection}:
\begin{equation}
    T_{\rm irr,p}^4 = \frac{\mu L_p}{4\pi \sigma_{\rm SB} r^2},
\end{equation}
but in this case $\mu$ is the flaring angle of the {surface of the disk (which is unknown and we set it to 0.1), $L_p$ the bolometric luminosity of the planet and $r$ the radial distance from the companion. All the values that we adopt are reported in the appendix, in Table \ref{tab:parameters}}.
Finally,
\begin{equation}
    T_{\rm acc}^4(r) = \frac{3GM_p \dot{M}_{\rm acc}}{8\pi \sigma_{\rm SB} r^3} \left(1 - \sqrt{\frac{r_p}{r}}\right)
\end{equation}
represents the contribution to the temperature profile given by accretion heating \citep{d1998accretion}. $M_p$ is the companion's mass, $\dot{M}_{\rm acc}$ its accretion rate and $r_p$ its radius.

\begin{figure}[h]
    \centering
    \includegraphics[width=1\columnwidth]{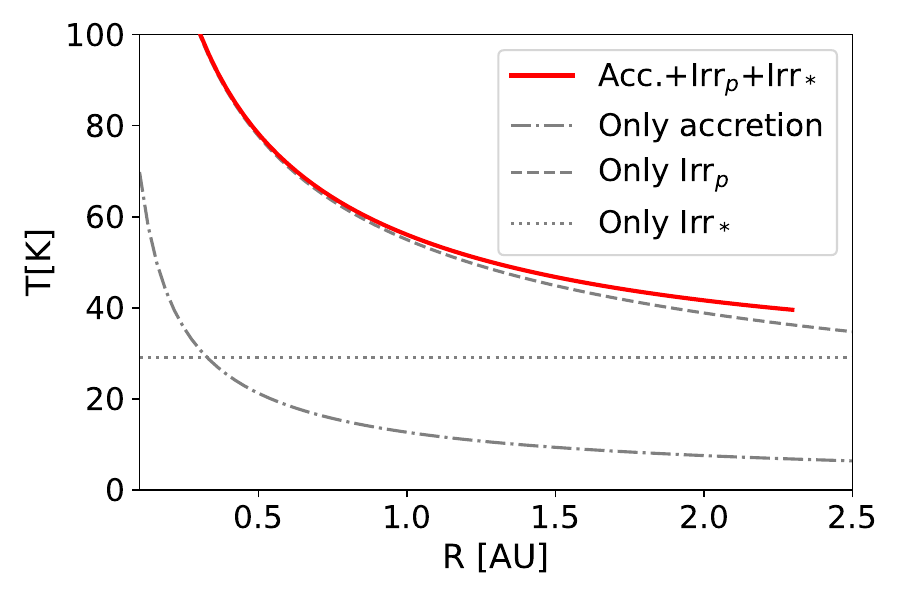}
    \caption{Different contributions to the temperature profile of GQ Lup B. The red line, which represents the total temperature profile, interrupts at the upper limit we retrieve from the beam size.}
    \label{fig:rTprofile}
\end{figure}

In the case of GQ Lup B, we find that for radii smaller than 2.3 au, the irradiation from the companion itself ($L=10^{-2.16}\,L_\odot $) is the dominant term (see Fig. \ref{fig:rTprofile}). The contribution of the accretion term to the global temperature profile is negligible, while the irradiation from the star, which effectively acts as a background, plays only a marginal role. We note that in our models this background term strongly depends on the grazing angle, which is unknown. While for GQ Lup B and SR 12 c this is almost irrelevant (we set it to $45^\circ$), the stellar irradiation term completely dominates the temperature profile of CPDs closer to the star and more accurate modeling should also take into account shadowing from the primary disk. 

Having a temperature profile, we are able to convert this to an emergent intensity assuming the full Planck law  ($B_\nu(T(r))$). The total flux $F_\nu$ can be obtained by integrating from a chosen inner radius $r_{in}$ to the outer radius $r_{out}$:
\begin{equation}
    F_{\nu} = \frac{2\pi \langle \cos i \rangle}{d^2} \int_{r_{\rm in}}^{r_{\rm out}} B_{\nu}(T(r)) \, r \, dr,
    \label{eq:totalflux}
\end{equation}
where $d$ is the distance of the system and $i$ is the disk inclination with respect to the line of sight. We find that the choice of the inner radius is not relevant for the final result, {as the dust temperature decreases relatively slowly over radius}.
Finally, we use Eq. (\ref{eq:totalflux}) to compute the outer radius which best reproduces the observed mm flux. All the parameters we used are listed in Appendix A in Table \ref{tab:parameters}. We obtain an outer radius of 0.20 au (418 R$_{\rm Jup}$), smaller than SR 12 c (0.34 au) and PDS 70 c (0.38 au, setting $\phi=1^\circ$).
As pointed out in \cite{wu2022alma} however, this value likely represents a lower limit: if dust emission was optically thin, it would scale with the dust mass and a more extended disk could have a similar brightness. 

\paragraph{Mass estimate from optically thin assumption}
While in the optically thick regime the flux density scales as the emitting surface, in the optically thin assumption it scales with the total mass. 
In the most simple scenario, at frequency $\nu$ the mass of optically thin dust is related to its flux density $F_\nu$ by \citep{1983QJRAS..24..267H}
\begin{equation}
    M_{\rm dust} = \frac{F_{\nu} \, d^2}{\kappa_{\nu} \, B_{\nu}(T_{\rm mean})},
    \label{eq:Mthin}
\end{equation}
where $\kappa_{\nu}$ is the dust opacity and $B_{\nu}(T_{\rm mean})$ is the Planck function evaluated at a mean representative temperature $T_{\rm mean}$.
For the dust opacity we assume mm-sized grains  ($\kappa_\nu=3.63 \,\rm cm ^2 \, \rm g ^{-1}$, \citealt{birnstiel2024dust}). We calculate a mean temperature of the disk of $\sim 56$ K by averaging the temperature profile (Eq. \ref{eq:temperature}) from a minimum radius of 4 R$_{\rm Jup}$ (a mean estimate of the companion's radius, \citealt{demars2023emission}) to the upper limit on the disk's radial extent, 2.3 au. 
We find a mass of 0.0032 $M_{\oplus}$, less than half of and a quarter of the analog estimates we obtain for PDS 70 c and SR 12 c (0.006 and 0.012 $M_{\oplus}$), respectively. Our estimates for PDS 70 c and SR 12 c agree with what is obtained in \cite{benisty2021circumplanetary} and \cite{wu2022alma}, but we note that these values, especially for PDS 70 c, strongly depend on the temperature profile, which is set by the choices on the parameters in Table \ref{tab:parameters}.

\paragraph{Accounting for the optical depth}

\begin{figure}[h]
    \centering
    \includegraphics[width=1\columnwidth]{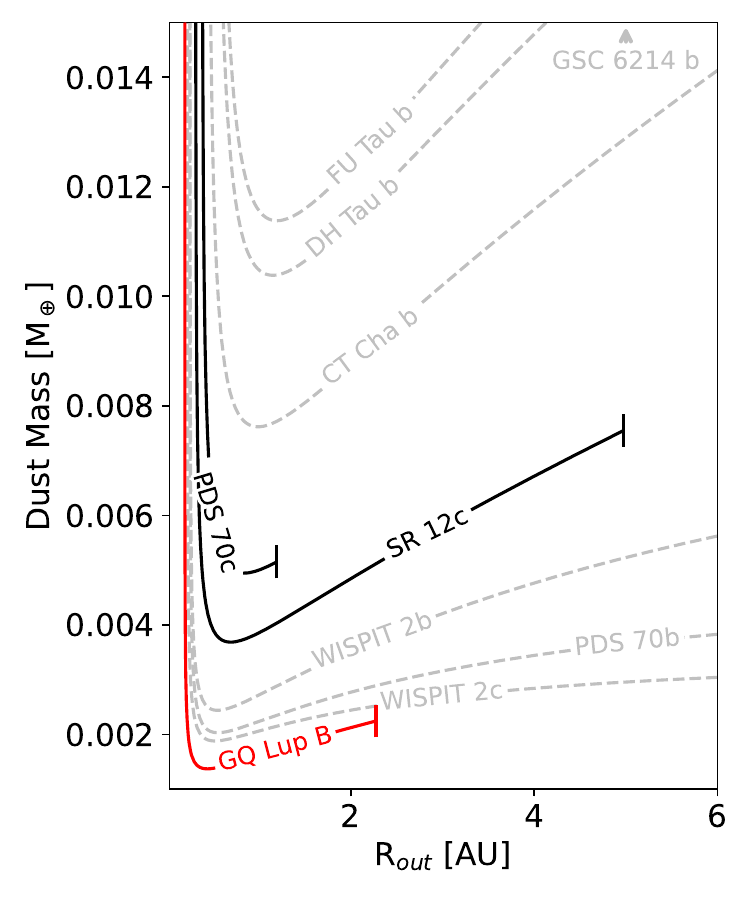}
    \caption{Continuous lines represent different combinations of {dust} masses and radial extents that can reproduce the observed millimeter flux density for GQ Lup B, SR 12 c and PDS 70 c. The lines stop with a vertical dash, representing the maximum radius estimated from the angular resolution of the observations. Dashed lines represent upper limits on other candidates \citep{bowler2015alma,wu2020alma, facchini20262, 2026ApJ..1009L..32B}. All the parameters are listed in Table \ref{tab:parameters}. }
    \label{fig:radiusvsmass}
\end{figure}

A more comprehensive approach to infer disk properties from a single submillimeter flux measurement is to assume a {dust} surface density profile $\Sigma(r)$ to account for the dust optical depth. Once again following \cite{isella2014searching,isella2019detection} and \cite{wu2022alma}, Eq. (\ref{eq:totalflux}) becomes 

\begin{equation}
    \begin{split}
    F_{\nu} = \frac{2\pi \cos i}{d^2} \int_{r_{in}}^{r_{out}} & \left[ 1 - \exp \left( -\frac{\Sigma(r) \kappa_{\nu}}{\cos i} \right) \right] \\
    & \times B_\nu [T(r)] r \, dr,
    \end{split}
\end{equation}
where we assume $\Sigma = \Sigma_0 r^{-1}$.
We then calculate the flux density for a grid of disk dust masses $M_d$ and outer radii $r_{out}$:
Fig. \ref{fig:radiusvsmass} shows the combinations of these parameters for which the calculated $F_\nu$ equals the observed peak intensity of GQ Lup B. We assume once again mm dust grains with opacity $\kappa_{\nu}$ = 3.63 $\text{cm}^2 ~\text{g}^{-1}$. 
In the same figure we show the same curve for PDS 70c and SR 12 c, together with dashed lines representing upper limits on other candidates {for disks around PMOs}. All of the parameters we used are listed in Table \ref{tab:parameters}.
In particular, at a first approximation we assume PDS 70 b and c and WISPIT 2 b and c to be aligned with the primary disk: we set the grazing angle to $1^\circ$, and for $i$ we use the primary disk's inclination. All the other objects (DH Tau b, GQ Lup B, FU Tau b, SR 12 c, CT Cha b) are wide orbit Planetary Mass Companions, and a priori there are no constraints on their disk's orientation. Hence, for them we set both angles to $45^\circ$. As highlighted by Fig. \ref{fig:rTprofile}, for PMCs the temperature profile is dominated by the companion's irradiation, and {the grazing angle of the primary's irradiation plays only a negligible role in the observed emission. Also the inclination with respect to the L.O.S. plays a marginal role, except for highly inclined disks ($i\gtrsim80^\circ$).}

Although coming from a relatively simple model, these curves highlight the different possible structure of the disks. 
A small source means high densities and optically thick dust: its emission is related only to the emitting surface, giving no constraints on the total mass. 
If the dust disk is instead larger, it quickly becomes optically thin, and the observed emission is related to its total mass. 
The curves behave differently at large radii because of the assumed temperature profile: emission is related to the temperature, which depends on the distance from the companion in different ways according to which term of Eq. (\ref{eq:temperature}) dominates. If the major component is the stellar irradiation, $T$ no longer depends on $r$, the curve becomes flat and we have a direct measure of the disk mass (as in PDS 70 b and WISPIT 2 b and c). 
However, if the stellar irradiation is low (as for SR 12 c), at large $r$ the temperature drops and larger masses are needed to reproduce the observed flux density. In this case inferring the disk mass is not trivial. 

While little can be said regarding the exact mass and radius of the dust disk around GQ Lup B, {in Fig. \ref{fig:radiusvsmass} its curve lies below those of SR 12 c and PDS 70 c, naively suggesting a smaller or less massive disk. This can be explained considering the relatively high irradiation from the companion (see the third column in Table ~\ref{tab:parameters}), }
which heats up the dust increasing the mm flux density. 

\subsection{Accretion rate}

\begin{figure*}[h]
    \sidecaption
    \includegraphics[width=12cm]{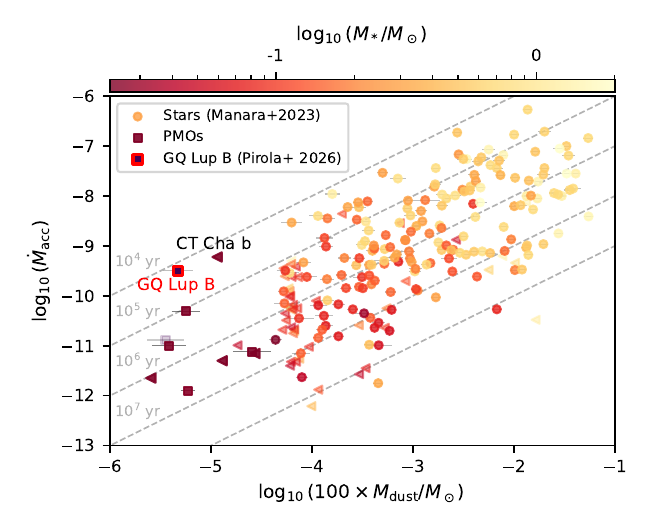}
    \caption{Accretion rates for disks around stars (circles) and planetary-mass objects (squares, the same discussed in Fig. \ref{fig:fvsmass}), against dust mass estimates (Eq \ref{eq:Mthin}, T=20 K) multiplied by 100 to account for the gas mass. Triangles represent upper limits. The color scale here accounts for the mass of the disk host, while dashed gray lines represent depletion timescales.}
    \label{fig:mdustvsmdot}
\end{figure*}

Figure \ref{fig:mdustvsmdot} compares accretion rates for planetary-mass objects with stars in nearby star-forming regions. Following \cite{manara2023demographics}, on the x-axis we plot an estimate of the total disk mass, using Eq. (\ref{eq:Mthin}) with $T=20$K to infer the dust mass and multiplying it by 100 to account for the gas reservoir, in the assumption that the gas-to-dust mass ratio is uniform across the sample.
Dashed gray lines indicate the survival timescale of a gas disk with the reported mass and accretion rate.
For stars, evolution models predict a correlation between the accretion rate and the global disk mass \citep{hartmann1998accretion,jones2012relationship,rosotti2017constraining}. 
Large samples from ALMA and spectroscopic surveys revealed that the relation is indeed typically distributed around ${M}_{\rm disk}/\dot{M}_{\rm acc} = 1 \, \rm Myr$ with significant scatter (Fig. \ref{fig:mdustvsmdot}), and does not appear to depend on stellar mass \citep{manara2023demographics}. 
Initial results suggest that brown dwarfs are systematically located in the lower part of the distribution \citep{sanchis2021measuring}, though the sample remains limited. 

The few planetary-mass objects for which mm detections are available (the same listed for Fig. \ref{fig:fvsmass}, reported as squares in both figures) appear broadly consistent with the stellar trend, but the limited sample and large scatter once again prevent any firm conclusion. 
We omit error bars on mass accretion rates ($\dot{M}_{\mathrm{acc}}$) because of their strong correlation with the models and parameters used: different estimates from the same measurement can vary up to an order of magnitude. 
Nonetheless, GQ Lup B appears on the high end of the scatter, showing a relatively high accretion rate (${\sim} 10^{-9.5}\,M_{\odot}\,\rm yr^{-1}$ \citealt{stolker2021characterizing}) for its mm-flux density. 
Notably, the mean temperature derived in the previous section ($\sim$ 56 K) is significantly higher than the 20 K assumed here: higher temperatures imply that the observed mm-flux density is compatible with even lower disk masses, which would shift the red dot in Fig. \ref{fig:mdustvsmdot} further to the left. 

Assuming the accretion rate to be constant in time, the gas component of the disk would be depleted in less than ${\sim} 10^{4}\,\rm yr$, two orders of magnitude shorter than the system's age \citep[3–5 Myr,][]{macgregor2017alma}. 
Two main observational caveats apply to this depletion timescale. 
First, the inferred $\dot{M}$ depends on the correction adopted for the extinction ($A_V \sim 2.3$ mag); previous work adopting lower estimates \citep{wu2015new}, inferred an accretion rate lower by an order of magnitude. 
Second, line emission studies show that accretion in GQ Lup B is likely variable \citep{stolker2021characterizing,demars2023emission}, implying that the time-averaged accretion rate could be lower than the single value shown here. 
Even accounting for these caveats, the depletion timescale assumes GQ Lup B to be isolated, which is not the case. Recent orbital studies \citep{venkatesan2025constraints} retrieved an orbital period of $\sim 1000$ yr and a highly inclined orbit ($\sim 63^\circ$ with respect to the primary disk) with a 97 au semimajor axis. Since the CO disk seems to extend up to 200 au \citep{macgregor2017alma}, the companion would puncture through the primary's gas disk twice per orbit, with a timescale of $\sim500$ yr. 
{This periodic interaction could continuously replenish the gas reservoir of the disk of GQ Lup B, naturally explaining the observed high accretion rate due to an increased inflow of material. However, the companion disk mass is inferred from the dust emission, whereas the PMO does not appear to cross the circumprimary mm-continuum disk. If the gas-to-dust ratio of the outer regions of the circumprimary disk is significantly above the factor of 100 assumed to produce Fig.~\ref{fig:mdustvsmdot}, the gas mass of the disk orbiting GQ Lup B may be significantly higher than anticipated.}

GQ Lup B is not the only planetary-mass companion with an accretion rate high enough to suggest gas depletion on timescales shorter than the system's age. CT Cha b \citep{wu2015new,cugno2025_ctcha} is also located in the upper region of Fig. \ref{fig:mdustvsmdot}: given the upper limit on its mm flux density, its disk mass could even be lower than that of GQ Lup B, making the implied accretion timescale even shorter and the tension with the companion's age even more pronounced. CT Cha b lies at a projected separation of $\sim$511 au from its host star, well outside the primary disk, meaning that its gas reservoir cannot be continuously replenished as in GQ Lup. This makes the persistence of an actively accreting disk even harder to explain, and suggests that additional mechanisms {as late infall} may be at play.
A thorough discussion of accretion models is beyond the scope of this paper. We note, however, that some theoretical frameworks predict episodic accretion \citep{baraffe2009episodic,wagner2018magellan,close2025three}, in which extended periods of near-quiescence are interrupted by short, intense bursts of accretion triggered by {gravitational and magnetorotational} instabilities. If such variability is common among PMCs, as seems plausible by recent findings \citep{2022NatAs...6..751C,close2025three, 2025ApJ...980L..39Z,2026AJ....172..134F}, instantaneous accretion rate measurements may not be representative of the time-averaged value, potentially alleviating the tension with the disk mass estimates.
We have also neglected the role of satellite formation, which can alter the disk structure and thus the observational signatures accessible to current facilities.
{To carry out our analysis, we had to make use of several assumptions (dust opacity, gas mass, inclination, radial extent, temperature), which often have no direct observational measurement. 
Multi-wavelength characterization of Planetary-Mass Companions is therefore crucial to robustly constrain properties of disks around PMOs.}

\section{Conclusions}

In this work we analyze ALMA continuum observations at 0.89 mm of the GQ Lup system, which hosts one of the best-known planetary-mass companions, GQ Lup B. Our main results are the following:
\begin{itemize}

\item We report the continuum detection of GQ Lup B. We retrieve a $5.8\sigma$ point source and, following the analysis carried out for the mid-IR excess, we associate it with the unresolved dust in a disk around the companion. This is the third millimeter detection of a gravitationally bound planetary-mass object, after PDS 70 c and SR 12 c.
\item The emission is co-spatial with both GRAVITY astrometry and SPHERE scattered light emission, and shows a peak intensity of $54.5\pm9.4\,\mu\rm Jy$. 
\item The source is unresolved at the best angular resolution we can achieve, implying a 2.3 au upper limit on the emission radius. 
\item In the optically thick approximation, we retrieve a dust disk radius of $0.2$ au, whereas in the optically thin regime we estimate a dust mass of $0.0032\,M_\oplus$.
\item GQ Lup B exhibits a remarkably high accretion rate for its millimeter flux density: this may be partly sustained by its eccentric orbit intersecting the primary's gas disk every $\sim$500 yr, replenishing the circumplanetary gas reservoir.
\end{itemize}

{Our millimeter detection completes an already rich dataset, including NIR and MIR spectral energy distributions, and upcoming JWST/NIRSpec $3-5~\mu$m spectra and JWST/MIRI mid-IR photometry. A natural next step would be to make use of this wealth of information in a thorough radiative transfer model, taking into account all the already available wavelengths. 
We also note that multi-frequency mm observations are needed to distinguish between different emission mechanisms (thermal emission from cold dust grains or free-free emission from accretion shocks). All of this is left for future works, while here we only focus on the 890 $\mu$m flux density we observe with ALMA, considering only dust thermal emission.}

The growing set of millimeter detections {of planetary-mass objects} is starting to provide a statistical overview, spanning a wide range of orbital configurations. 
In this frame, PMCs are emerging as a crucial laboratory to investigate {the physics of disks around substellar objects, bridging circumstellar and circumplanetary environments.} Our detection demonstrates the sensitivity required to detect these faint structures at submillimeter wavelengths: enlarging the current sample will be crucial to assess whether disks around substellar objects are similar to scaled-down versions of circumstellar disks, or additional physical processes govern their formation and evolution.

\begin{acknowledgements}

This paper makes use of the following ALMA data: ADS/JAO.ALMA 2023.1.01340.S, 2024.1.00780.S. ALMA is a partnership of ESO (representing its member states), NSF (USA) and NINS (Japan), together with NRC (Canada), NSTC and ASIAA (Taiwan), and KASI (Republic of Korea), in cooperation with the Republic of Chile. The Joint ALMA Observatory is operated by ESO, AUI/NRAO and NAOJ. 
S.F. acknowledges financial contribution from the European Union (ERC, UNVEIL, 101076613). Views and opinions expressed are however those of the authors only and do not necessarily reflect those of the European Union or the European Research Council. Neither the European Union nor the granting authority can be held responsible for them.
P.C. acknowledges support by the ANID BASAL project FB210003.
J.B. is supported in part by JWST Cycle 3 Program JWST-GO-06739 provided by NASA through a grant from the Space Telescope Science Institute, which is operated by the Association of Universities for Research in Astronomy, Inc., under NASA contract NAS5-03127. 
M.B. has received funding from the European Research Council (ERC) under the European Union’s Horizon 2020 research and innovation programme (PROTOPLANETS, grant agreement No. 101002188).
\end{acknowledgements}

\bibliographystyle{aa}
\bibliography{bibliography}

\onecolumn
\begin{appendix}

\section{Imaging varying robust}
\begin{figure}[h]
    \centering
    \includegraphics[width=1\textwidth]{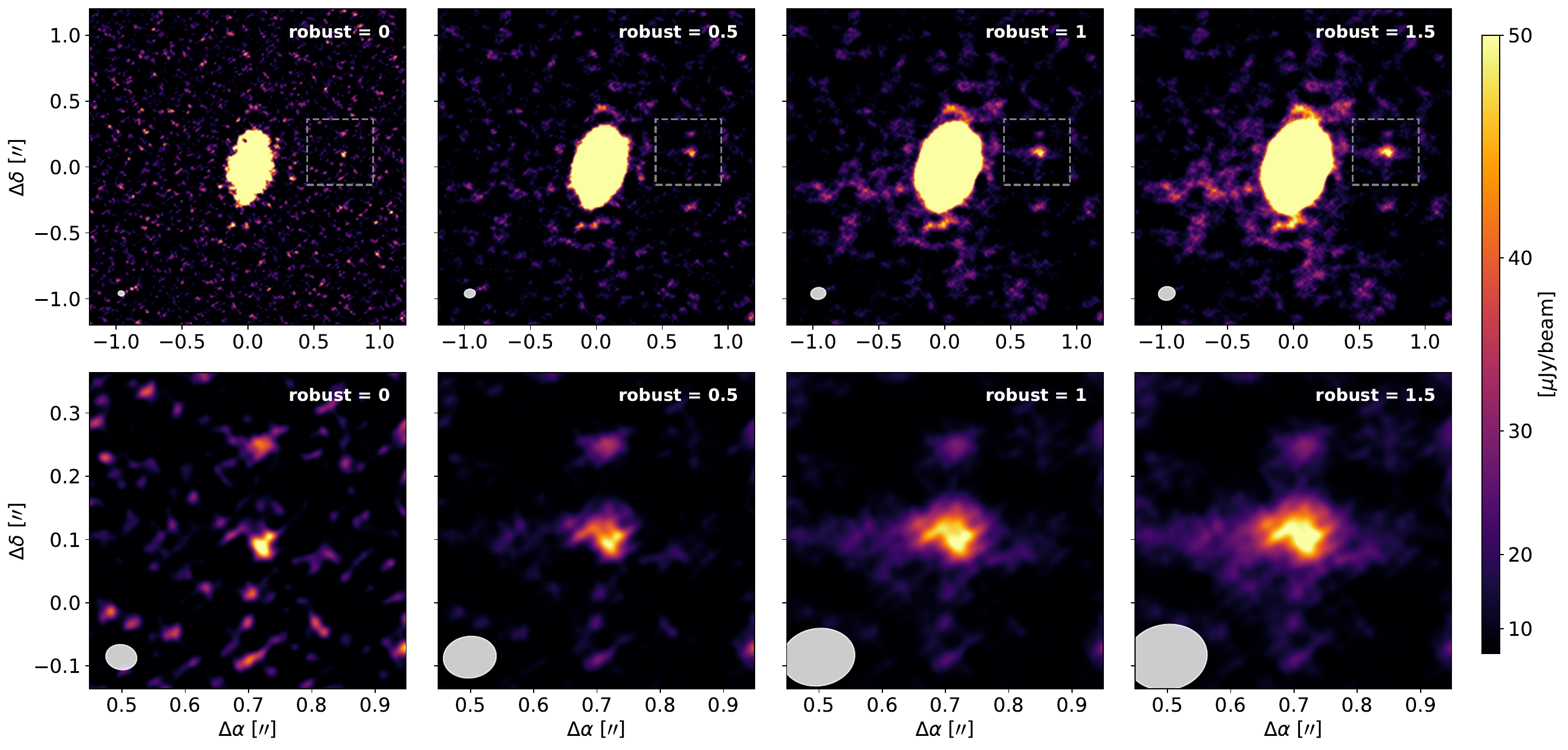}
    \caption{Images obtained with different Briggs robust parameters, which translate into different
    beam sizes. The bottom row represents the zoomed-in area around the companion, indicated with a gray square in the top row. The color scale is the same as in Fig. \ref{fig:continuumvsscattered}.}
    \label{fig:varying_robust_combined}
\end{figure}

\section{Residuals of imaging varying threshold, masking only the primary disk}
\begin{figure}[h]
    \centering
    \includegraphics[width=1\textwidth]{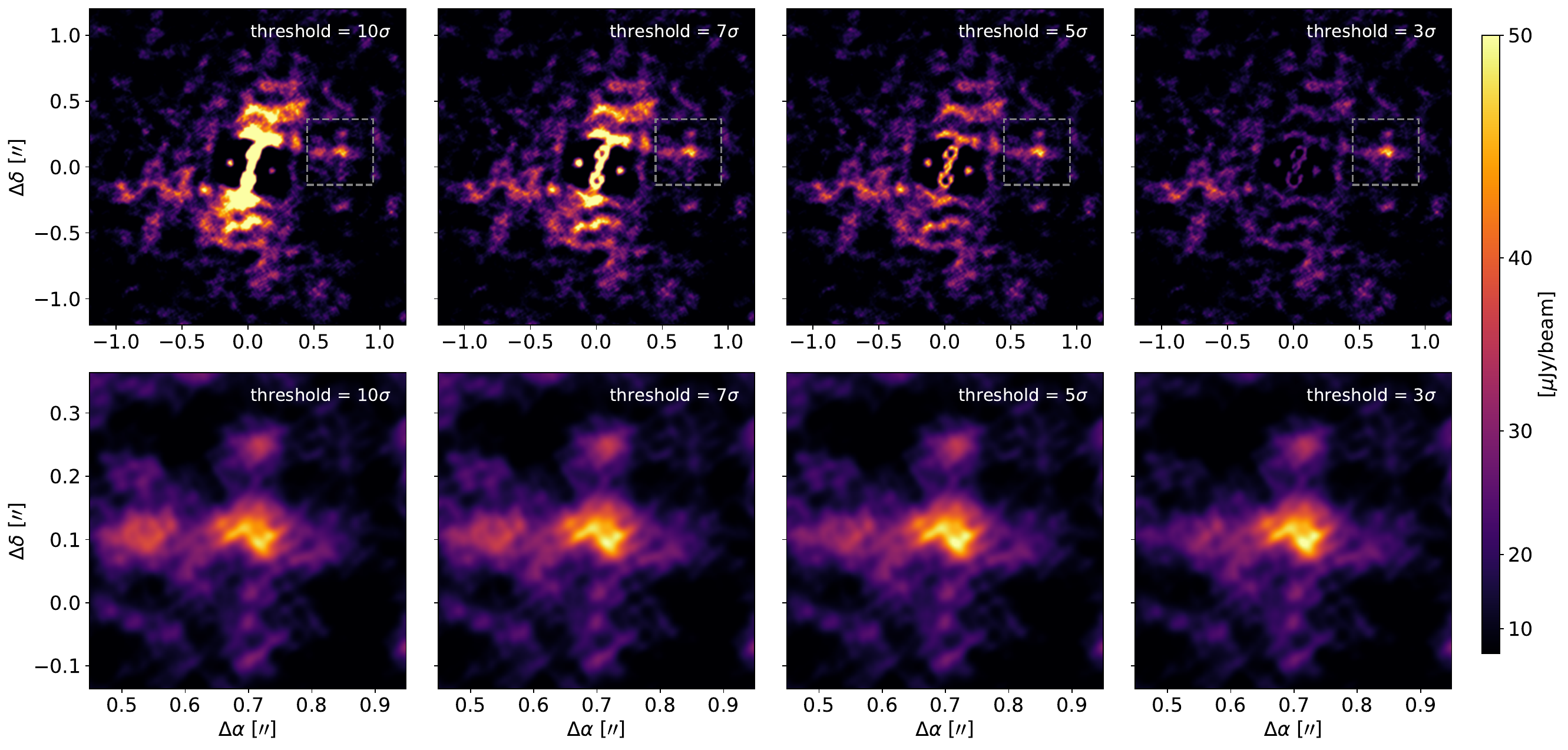}
    \caption{Residual map of the images obtained masking only the disk of the primary star and stopping the imaging
    procedure at different thresholds. The bottom row represents the zoomed-in area around the companion, indicated with a gray square in the top row. The color scale is the same as in Fig. \ref{fig:continuumvsscattered}, and the maps are the same used for Fig. \ref{fig:test2}.}
    \label{fig:varying_thresh_combined}
\end{figure}

\section{Radial profile comparison}
\begin{figure}[h]
    \sidecaption
    \includegraphics[width=12cm]{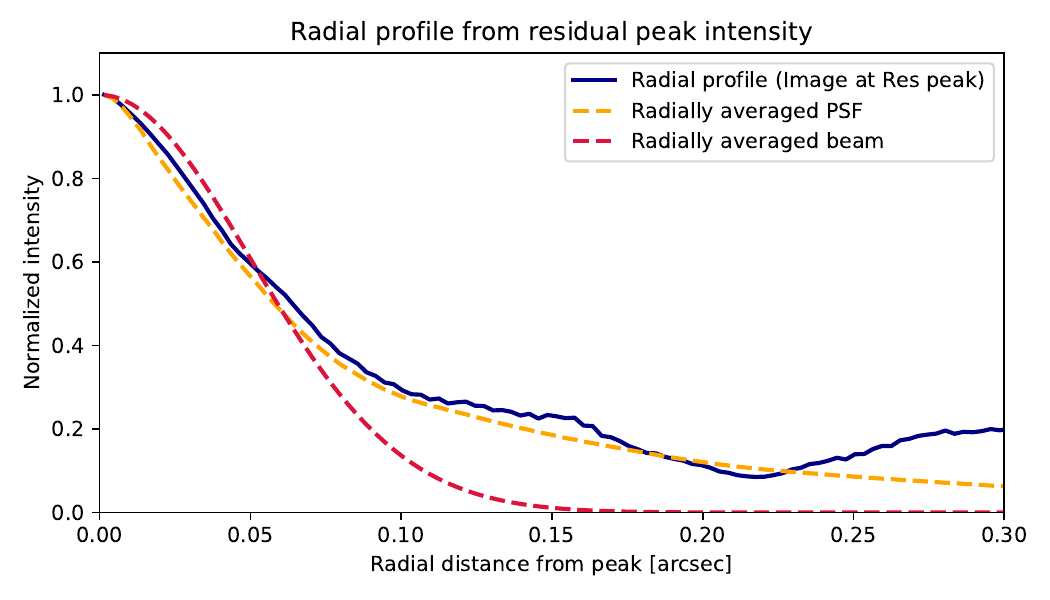}
    \caption{Azimuthal average of the clean beam and the PSF of the image obtained with robust = 2 (Fig. \ref{fig:continuum_fullscale}), compared with the azimuthally averaged profile of the emission, centered in the position of the companion (taken as the position of the peak of the residual map). The radial profile broadly follows the PSF profile, coherently with an unresolved point source slightly above the cleaning threshold.}
    \label{fig:beam_radial_profiles}
\end{figure}

\begin{table}[h]
\section{Parameters used in modeling}
\caption{Parameters used to calculate the disk temperature profiles.}
\label{tab:parameters}
\centering
\begin{tabular*}{\textwidth}{@{\extracolsep{\fill}}lccccccc}
\hline\hline
PMC & $M_p$ & $\log_{10}(L_p/L_\odot)$ & $\dot{M}_{\rm acc}$ & $L_*$ & $D$ & $d$ & $F_{\rm 345\,GHz}$ \\
 & ($M_{\rm Jup}$) & & ($M_\odot$\,yr$^{-1}$) & ($L_\odot$) & (pc) & (au) & (mJy) \\
\hline
GQ~Lup~B & 20~[1] & $-2.16$~[2] & $10^{-9.5}$~[1] & 0.8~[3] & 154~[4] & 97~[5] & 0.054 \\
PDS~70~b & 4~[6] & $-3.80$~[7] & $1.30\times10^{-11}$~[8] & 0.36~[7] & 112~[4] & 22~[9] & 0.080~[10] \\
PDS~70~c & 7~[6] & $-3.80$~[7] & $10^{-11}$~[8] & 0.36~[7] & 112~[4] & 34~[9] & 0.086~[10] \\
SR~12~c & 16~[11] & $-2.95$~[12] & $10^{-11}$~[12] & 0.12~[12] & 112.5~[4] & 1200~[11] & 0.127~[12] \\
WISPIT~2~b & 4.9~[13] & $-3.5$~[14] & $2.25\times10^{-12}$~[14] & 0.7~[14] & 133~[4] & 57~[14] & $<0.040$~[15] \\
WISPIT~2~c & 9~[16] & $-3.6$~[16] & $< 10^{-13}$~[16] & 0.7~[14] & 133~[4] & 15~[16] & $<0.040$~[15] \\
FU~Tau~b & 19~[17] & $-2.51$~[17] & $6\times10^{-11}$~[17] & 0.04~[18] & 126~[4] & 726~[17] & $<0.200$~[19] \\
DH~Tau~b & 14~[19] & $-2.72$~[20] & $5.01\times10^{-12}$~[20] & 0.27~[19] & 135~[4] & 330~[19] & $<0.200$~[19] \\
CT~Cha~b & 17~[19] & $-2.62$~[19] & $6.00\times10^{-10}$~[21] & 0.57~[19] & 190~[4] & 440~[19] & $<0.090$~[19] \\
GSC~6214-210~b & 14~[22] & $-2.8$~[22] & $7.00\times10^{-12}$~[20] & 0.38~[22] & 145~[4] & 330~[22] & $<0.367$~[22] \\
\hline
\end{tabular*}
\tablefoot{The original measurements with the relative uncertainties can be found in the references listed below.}
\tablebib{
[1]~\citet{stolker2021characterizing}; [2]~\citet{marois2007gq}; [3]~\citet{demars2023emission};
[4]~\citet{collaboration2022gaia}; [5]~\citet{venkatesan2025constraints};
[6]~\citet{wang2021constraining}, \citet{trevascus2025differentiating}; [7]~\citet{muller2018orbital};
[8]~\citet{haffert2019two}; [9]~\citet{mesa2019vlt}; [10]~\citet{benisty2021circumplanetary};
[11]~\citet{2026AJ....172..134F}; [12]~\citet{wu2022alma}; [13]~\citet{van2025wide};
[14]~\citet{close2025wide}; [15]~\citet{facchini20262}; [16]~\citet{lawlor2026direct};
[17]~\citet{wu2023monitoring}; [18]~\citet{bailer2021estimating}; [19]~\citet{wu2020alma} (Table 2);
[20]~\citet{zhou2014accretion}; [21]~\citet{wu2015new}; [22]~\citet{bowler2015alma}.
}
\end{table}

\FloatBarrier 
\twocolumn
\end{appendix}

\end{document}